\documentclass[aps,prx,twocolumn,showpacs,superscriptaddress,groupedaddress]{revtex4-1} \pdfoutput=1
\usepackage{graphicx} 
\usepackage{amsmath}
\usepackage{graphicx}
\usepackage{natbib}

\usepackage{amsthm, amsmath, amssymb}
\usepackage[loose,nice]{units} 

   \def\a{\alpha}

\def\p{\mathbf{p}}

\begin{document}
\title{Rotational random walk of the harmonic three body system}
\date{\today}
\author{Ori Katz-Saporta} \affiliation{Department of Physics of Complex Systems,
Weizmann Institute of Science, Rehovot 76100, Isreal} \author{Efi
Efrati} \email{efi.efrati@weizmann.ac.il} \affiliation{Department of
Physics of Complex Systems, Weizmann Institute of Science, Rehovot
76100, Isreal}

%

\begin{abstract} 
When Robert Brown first observed colloidal pollen grains in water he inaccurately concluded that their motion arose "neither from currents in the fluid, nor from its gradual evaporation, but belonged to the particle itself". In this work we study the dynamics of a classical molecule consisting of three masses and three harmonic springs in free space that does display a rotational random walk "belonging to the particle itself". The geometric nonlinearities arising from the non-zero rest lengths of the springs connecting the masses break the integrability of the harmonic system and lead to chaotic dynamics in many regimes of phase space. The non-trivial connection of the system's shape space allows it, much like falling cats, to rotate with zero angular momentum and manifest its chaotic dynamics as an orientational random walk. In the transition to chaos the system displays random orientation reversals and  provides a simple realization of  L{\'{e}}vy walks.
\end{abstract} \maketitle
A cat in free-fall manages to rotate mid-air and land with its feet on the ground even when dropped from rest and having zero angular momentum throughout its fall. This type of motion is counterintuitive as our intuition often relies on the mechanics of rigid bodies where any finite-rate rotation is necessarily associated with angular momentum. However, a cat is not a rigid body. Its ability to deform allows it to manuever into a new orientation with zero overall angular momentum \cite{Guichardet1984}. The cat performs a cyclic series of internal deformations at the end of which it returns to its initial shape,
reoriented in space. Such motions, which we term deformation-induced rotations, have been studied extensively in the context of molecular motions \cite{Guichardet1984}, falling cats \cite{Montgomery1993}, gymnastic maneuvers \cite{Frohlich1980} and stellar systems \cite{Montgomery2015}. In these systems the orientation of the object considered is neither a free variable nor fully determined by the values of the other independent variables describing the system. The dynamics of the orientation can be expressed through the dynamics of the other independent variables in the system, yet to obtain the orientation of the system at a given point in time one needs to know the full history of the dynamics of the other independent variables. Such systems, which are called non-holonomic \cite{Batterman2003}, arise in a wide variety of settings ranging from the physics of parallel parking \cite{Paromtchik1996}, to robotic motion \cite{KellyMurray1995}, to autonomous motion in curved space \cite{Wisdom2003}. The dynamics of these systems are often formulated in terms of a gauge field on the space of free variables \cite{Batterman2003}. The history dependence of the evolution of the non-holomonically constrained variables manifests in the cumulation of a relevant geometric phase, reminiscent of Berry's phase \cite{wilczek1989geometric,Ber84}, along a given trajectory of the system. In the present case it allows the orientation of the system to serve as a sensitive measurable of the nature of the dynamics of the deforming system.  In particular it allows the orientation to manifest the distinct cumulative characteristics of periodic, quasi-periodic and chaotic dynamics. 

\begin{figure}[ht]
\centerline{\includegraphics[width=.45\textwidth]{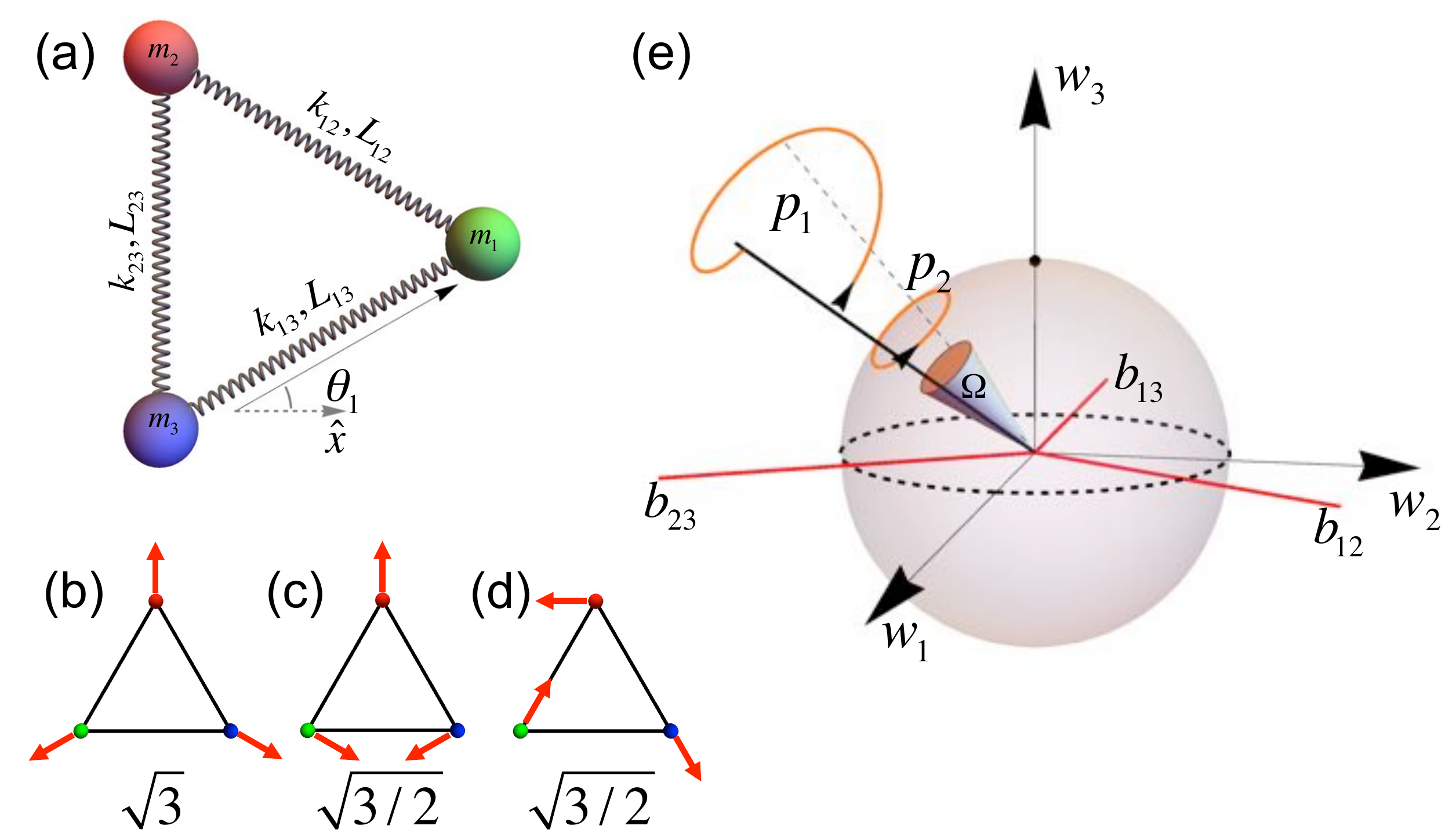}}
\caption{The harmonic three body system and its shape sphere. (a)
An illustration of the harmonic three body system, a triatomic harmonic spring-mass system. $\theta_1$ is the angle between $r_{12}$ and the $x$-axis, and is used to determine the orientation of the triangle in the plane. (b), (c) and (d) present the normal modes of the system when expanded to leading order around its zero-energy configuration; they are commonly known as the symmetric stretch, isometric bend and asymmetric bend, respectively. (e) The shape sphere. Every shape of the triangular spring mass system corresponds to a unique value of the parameters $w_{1}$, $w_{2}$ and $w_{3}$, introduced in \cite{Iwai1987} by Iwai, see relation to Jacobi coordinates in Supplementary Information (SI). All triangular shapes related by similarities are located on radial rays, and thus the unit sphere represents the unscaled triangle's shape space. When a triangle follows a path in which the initial and final points are similar triangles, for example $p_1$ in the figure, the path forms a close curve on the unit shape sphere, $p_2$ in the example. One can show that in this case the triangle's rotation  is given by the solid angle enclosed by the path on the unit sphere;
 $\Delta \theta_{1}=\Omega$.}
\label{fig:spring-mass}
\end{figure}

\begin{figure*}[ht]
\centerline{\includegraphics[width=.95\textwidth]{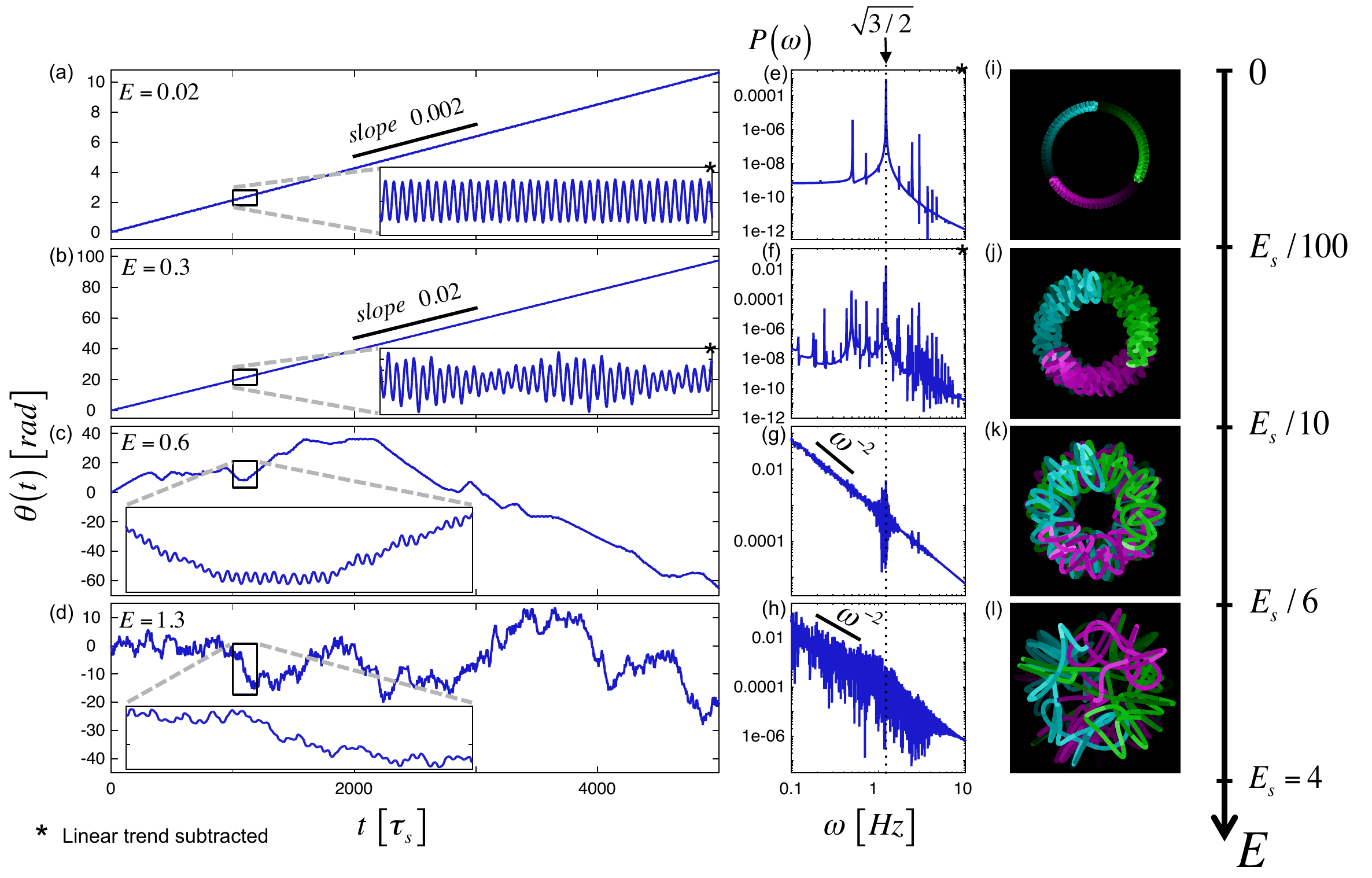}}
\caption{Four typical trajectories of the orientation of the three mass
triangle as a function of time, (a,b,c,d), their power spectra (e,f,g,h)
and a long-exposure image of their dynamics with each mass colored
in a different color (i,j,k,l). The initial conditions are determined
by a choice of the normal mode intensities of the reduced system (as explained
in the main text), and are chosen as: (a,e,i) $E=0.02$; (b,f,j)
$E=0.28$; (c,g,k) $E=0.62$; (d,h,l) $E=1.29$. The energy units
fit the simulation parameters $m=1$, $L=2$, $k=1$ and are compared
to the energy scale $E_{s}=3kL^{2}/2=6$ which corresponds to the energy needed to contract the triangle to a point.
(a,e,i) At low energies $E\ll E_{s}$ the system behaves quasi-periodically,
with frequencies fitting the linear modes of the system as calculated
analytically, peaking at the degenerate linear frequency $\sqrt{3/2}$.
(b,f,j) At slightly higher energies $E_{s}/100<E<E_{s}/15$ the degenerate
frequency $\sqrt{3/2}$ splits and a beating phenomenon is observed,
signifying a non-linear effect in the periodic regime.
(c,g,k) At higher energies $E_{s}/15<E<E_{s}/9$ the system enters
a chaotic regime, exhibiting periodic normal-mode motion on short
time scales, and seemingly random transition between constant angular
velocity bouts on long time scales. The power spectrum fills up, adopting a constant $\omega^{-2}$ slope fitting Brownian motion yet continues to display a significant peak around the linear frequency $\sqrt{3/2}$.
(d,h,l) Around the energy $E>E_{s}/9$ the system loses its short-time
periodicity and portrays seemingly random behavior. The power spectrum
shows that all frequencies are excited in this regime with approximately
the same power, and the linear frequencies lose  significance.}
\label{fig:trajectories}
\end{figure*}

For a dynamical system to display random walk characteristics its dynamics must loose correlations. This can be achieved in an isolated Hamiltonian system provided it is not integrable. The Hamiltonian dynamics of two point particles interacting only with each other via a central force is integrable regardless of the interaction potential \footnote{The two body problem for point masses can be shown to be planar. For a planar Hamiltonian system of two particles we seek four constants of motion. The invariants $H,\,p_x,\, p_y$ and $L$ are in involution and thus the system is completely integrable.}. However, three interacting particles suffice to yield chaotic dynamics as is well known from studying the classical gravitational three body problem. Arguably, the simplest interaction such particles can exhibit is a pairwise harmonic central force, quadratic in the distance between each two particles, i.e. connecting every two particles by a linear spring. If the springs' rest lengths are zero the problem can be shown to be integrable \cite{CS93}. However, the more natural system with finite spring rest lengths leads to richer dynamics. In this case the Hamiltonian reads 
\begin{equation} 
\begin{aligned}
\mathcal{H}=&\sum_{i=1}^{3}\frac{\p_{i}^{2}}{2m_{i}}+\sum_{<ij>}\frac{k_{ij}}{2}\left
(r_{ij}-L_{ij}\right)^{2}, 
\label{eq:hamiltonian} 
\end{aligned}
\end{equation} 
where, $\mathbf{r}_{ij}=\mathbf{r}_{i}-\mathbf{r}_{j}$, and $r_{ij}=|\mathbf{r}_{ij}|\equiv\sqrt{\mathbf{r}_{ij}\cdot\mathbf{r}_{ij}}$. 
If the total angular momentum of the system is zero, its motion is
constrained to a plane \cite{Whittaker1988}; it thus suffices to
analyze the problem in two dimensions. While every pair-wise interaction is purely harmonic, the
geometric non-linearity originating from the non-vanishing rest lengths
of the springs breaks the integrability of the system. These geometric
non-linearities manifest algebraically in  the  square-root in the
potential term of the Hamiltonian. We note that the three body system is also the simplest capable of displaying zero angular momentum rotations \footnote{for the two body system the only internal shape variable is the distance between the two particles, and thus every closed trajectory is necessarily reciprocal and does not lead to rotation \cite{Purcell1977}}. 

We now come to study the dynamics of the harmonic three body system by numerically integrating the symplectic dynamics of the Hamiltonian \eqref{eq:hamiltonian}; see Supplementary Information (SI) for further details. The system's parameters are chosen to be uniform for all the masses and springs; $m_{i}=1$, $k_{ij}=1$ and $L_{ij}=2$, and the corresponding equilibrium state of the system is an equilateral triangle. The typical time scale is $\tau\sim\sqrt{m/k}=1$ and the corresponding typical energy scale is $\mathcal{E}\sim3/2 k L^{2}=6$, the elastic energy required to double the size of the triangle or equivalently shrink it to a point. All the initial conditions prescribed to the system have zero angular and linear momentum. The time evolution of the orientation of the triangle for four typical numerically calculated trajectories is depicted in Figure \ref{fig:trajectories}.

Initial conditions of very small energy result in regular oscillatory motion superimposed on rotation with a constant average angular velocity (with zero angular momentum). In this regime the normal modes of motion are harmonic and non-interacting. As the energy of the system is slightly increased the modes begin to interact, their frequencies shift and the oscillations deviate from the harmonic form. However, the overall dynamics of the orientation remains that of oscillatory motion superimposed on a constant averaged angular velocity. Further increasing the energy content of the system breaks the regularity of its motion. For short times the angle dynamics is reminiscent of the dynamics observed in the regular regime exhibiting a constant averaged angular velocity motion. However, this motion does not persist indefinitely; instead after a finite time the orientation of the constant angular velocity reverses and, for example, a counter-clockwise rotating system will start rotating clockwise. The magnitude of the average angular velocity remains relatively constant, and this motion will again persist for a finite time before inverting again. Over long periods of time this dynamics leads to angular L{\'{e}}vy-flight statistics. The lengths of the bouts of time the angular velocity persists in a given direction shorten as the energy is further increased. For sufficiently high energy this leads to an angular random walk where the mean squared angular displacement grows linearly with time.

Focusing on infinitesimal perturbations of the equilibrium equilateral triangle we may na{\"i}vely linearize the Hamiltonian about a particular rest configuration of the spring-mass triangle, see SI. This textbook-type exercise \cite{Kotkin1971} yields frequencies that agree well with the frequencies observed in Figure \ref{fig:trajectories}.b. Yet, it describes oscillations about a specific state and fails to produce the constant average-angular-velocity motion observed. In order to properly capture the motion of the spring-mass system we employ a reduced description that does not restrict the dynamics of the system. The finite rest lengths of the springs provide the spring-mass system with a natural triangular shape that is preserved for small enough energy excitations, and about which the system's shape oscillates. We thus seek a description of the system as a deforming triangle placed in the plane instead of as three masses moving independently. The six degrees of freedom characterizing the $x$ and $y$ coordinates of the three point masses can be replaced by three variables describing the shape of the triangle, for example its side lengths, and three variables describing its position and orientation in space. To ease the subsequent analysis we employ the shape sphere variables used in \cite{Iwai1987} to describe the shape of the triangle:
\begin{align} 
w_{1} &=\frac{m}{6}\left(|\mathbf{r}_{23}|^{2}+|\mathbf{r}_{13}|^{2}-4\mathbf{r}_{23}\cdot\mathbf{r}_{13}\right),
 \nonumber \\ 
 w_{2} &=\frac{m}{2\sqrt{3}}\left(|\mathbf{r}_{23}|^{2}-|\mathbf{r}_{13}|^{2} \right), \nonumber \\
 w_{3} &= \frac{m}{\sqrt{3}}\left(\mathbf{r}_{13}\wedge \mathbf{r}_{23}\right).
 \label{eq:jacobi} \end{align}
Note that the above variables are invariant under rigid motions of the triangle, and that $w_{3}$ changes sign under reflection; see SI for the general derivation of these special coordinates and their interpretation. To fully describe the system three additional variables are required; we choose the center of mass coordinates, $\mathbf{R}_{cm}=(\mathbf{r}_{1}+\mathbf{r}_{2}+\mathbf{r}_{3})/3$, and the angle $\theta_{1}$ formed between $\mathbf{r}_{12}$ and the $x$-axis, as appears in Figure \ref{fig:spring-mass}. From these six variables the original six variables determining the two dimensional location of the three masses can be recovered uniquely.  $\mathbf{R}_{cm}$ and $\theta_{1}$ are cyclic coordinates and only their time derivatives appear in the Hamiltonian. The conservation of linear momentum poses a holonomic constraint on the problem; knowledge of $\mathbf{R}_{cm}(t=0)$ and  $\dot{\mathbf{R}}_{cm}(t=0)$ allows explicit determination of $\mathbf{R}_{cm}(t)$ for all $t$. The conservation of angular momentum, however, is non-holonomic. It leads to the following equation for the orientation evolution:
\begin{equation}
\dot{\theta}_{1}=\frac{w_{2}\dot{w}_{3}-w_{3}\dot{w}_{2}}{2w\left(w+w_{1}\right)}+\frac{L}{2w}
\label{eq:theta1},
\end{equation}
where $L$ is the angular momentum and $w=|\mathbf{w}|$.  
In contrast with the conservation of linear momentum, here knowledge of the initial value $\theta_{1}(t=0)$, the angular momentum $L$ and the variables $\mathbf{w}(t)$ is not sufficient to determine $\theta_{1}(t)$. Instead the full history of the $\mathbf{w}(t)$ variables evolution is required. The simplest way to demonstrate this is to follow a closed path in the independent coordinate base and show that in the general case such a trajectory leads to a non-trivial angle change 
$\Delta \theta_{1}$. As we are only interested in the zero angular momentum case we set $L=0$. In this case the calculation of the angle change along a closed path becomes a purely geometric problem. To elucidate this we use the polar decomposition of the $w_{i}$ coordinates
\[
\mathbf{w}=
(w_{1},w_{2},w_{3})=w(\sin\chi\cos\psi,\sin\chi\sin\psi,\cos\chi),
\]
and integrate equation \eqref{eq:theta1} along the closed path $\mathbf{w}\left(t\right)$. Making use of Green's theorem allows reducing $\Delta\theta_1$ to the solid angle the path encloses on the unit sphere, clarifying its nature as a geometric phase \cite{Montgomery2015}: 
\begin{align*}
\theta_{1}\left(t\right)-\theta_{1}\left(0\right)  &=\int_{\vec{w}\left(t\right)}\frac{-\sin\psi d\chi-\cos\psi\cos\chi\sin\chi d\psi}{2+2\cos\chi\cos\psi}\\ 
&=-\frac{1}{2}\iint\sin\chi\left(t\right)d\chi d\psi=-\frac{1}{2}\iint d\Omega,
\end{align*}
We note that as $\dot\theta_{1}$ does not depend on $w$ only the projection of the path to the unit sphere needs to be closed.
The end points of such paths lay on a radial ray representing similar triangles, and thus naturally allow comparing their orientations and rendering the calculated rotation $\Delta \theta_1$ gauge invariant (see Figure \ref{fig:spring-mass}e).

Substituting equation \eqref{eq:theta1} into the Hamiltonian yields a reduced Hamiltonian that depends only on the three $w_{i}$ coordinates and their conjugate momenta $p_{i}$
\begin{equation}
H_{red}   = w\left(p_{1}^{2}+p_{2}^{2}+p_{3}^{2}\right)+
\sum_{<ij>}\frac{k}{2}\left(r_{ij}(\vec{w})-L\right)^{2}
\end{equation} 
where
\[
\begin{aligned}
p_{i}=&\frac{\dot{w}_{i}}{2w},\qquad r_{ij}\left(\vec{w}\right)=\sqrt{2\left(w-\vec{w}\cdot\vec{b}_{ij}\right)}\, ,\\
\vec{b}_{13}=&\left(\tfrac{1}{2},\tfrac{\sqrt{3}}{2},0\right),
\,\,
\vec{b}_{12}=\left(-1,0,0\right),\,\,
\vec{b}_{23}=\left(\tfrac{1}{2},-\tfrac{\sqrt{3}}{2},0\right).
\label{eq:bij}
\end{aligned}
\]
This symplectic reduction and the geometric interpretation of the quantities that appear in it are provided in 
\cite{Iwai1987,Montgomery2015,Littlejohn1997,Marsden2000} and in the supplementary material. We note again that while the reduced Hamiltonian contains only degrees of freedom that relate to the shape of the triangle, in order to recover the full dynamics of the triangle we are required to integrate equation \ref{eq:theta1} to determine the temporal evolution of $\theta_1$.

We can now return to examine the small energy limit without restricting the motion of the masses. We consider small perturbations about the equilibrium shape of the Hamiltonian:
$\vec{w}=(0,0,2)+\epsilon\, \vec{\delta w}, \quad \vec{p}=\epsilon\, \vec{\delta p}$. While the coordinates $w_{1}, w_{2}$ and $w_{3}$ were chosen to provide the most transparent geometric characterization of the shape space of the deforming triangle, they also reduce the perturbed Hamiltonian to its normal form \cite{ArnoldDynamical1988}
\[
H_{red}\approx \epsilon^{2} \bigl[ 2\left(\delta p_{1}^{2}+\delta p_{2}^{2}+\delta p_{3}^{2}\right)+
3/16\left(\delta w_{1}^{2}+\delta w_{2}^{2}+2\delta w_{3}^{2}\right) \bigr].
\]
Note that the frequencies corresponding to the first two normal modes are degenerate; $\omega_{1}=\omega_{2}=\sqrt{3}/2$, and $\omega_{3}=\sqrt{3}$.

\begin{figure}[ht]
\centerline{\includegraphics[width=.5\textwidth]{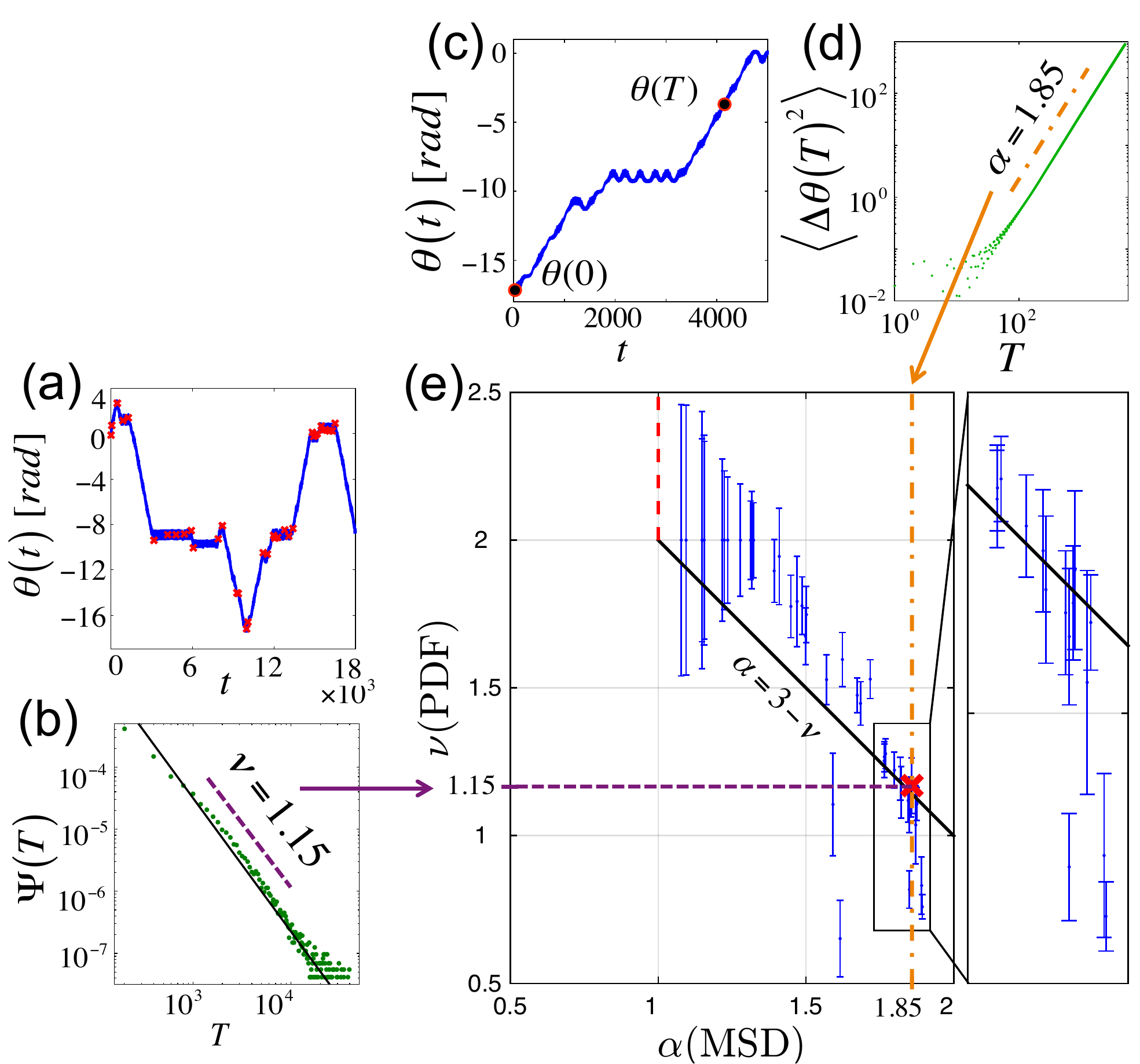}}
\caption{Comparison between the angular MSD and the bout-length
PDF in the regime of motion $E_{s}/15<E<E_{s}/9$ where the motion
is composed from bouts of constant average angular velocity of varying
times.
(a,b) show the process of calculating the bout-length PDF $\Psi\left(T\right)$,
by identifying the turning points between segments of overall constant
angular velocity. (a) shows a given trajectory with red asteri marking
the turning points identified by the algorithm. (b) shows the resulting
$\Psi\left(T\right)$ in a log-log plot along with the best linear
fit.
(c,d) show the process of calculating the MSD, by averaging $\left(\theta\left(T\right)-\theta\left(0\right)\right)^{2}$
over $10^{4}$ runs (see SI).
When $T$ is big enough, the MSD is an excellent fit to a power law
$\left\langle \Delta\theta\left(T\right)^{2}\right\rangle \propto T^{\alpha}$
with an anomalous diffusion exponent $\alpha$, as depicted in (d).
(e) The PDF power $\nu$ as a function of the anomalous diffusion
exponent $\alpha$ for several initial conditions in the relevant
energy range $E_{s}/15<E<E_{s}/9$. The black solid line and red dotted line show the
L{\'{e}}vy-walk model prediction $\alpha=3-\nu$. For low energy (the zoomed-in region) the two exponents follow the 
L{\'{e}}vy-walk prediction well.  As the energy grows, the average bout length shortens
and $\alpha$ decreases, changing from almost $2$ to $1$ at high
enough energies. The PDF fit  to a single power law gradually deteriorates with increasing energy and we observe significant deviations from the 
power-law based L{\'{e}}vy-walk prediction. 
}
\label{fig:Levy}
\end{figure}

\begin{figure*}[ht]
\centerline{\includegraphics[width=.85\textwidth]{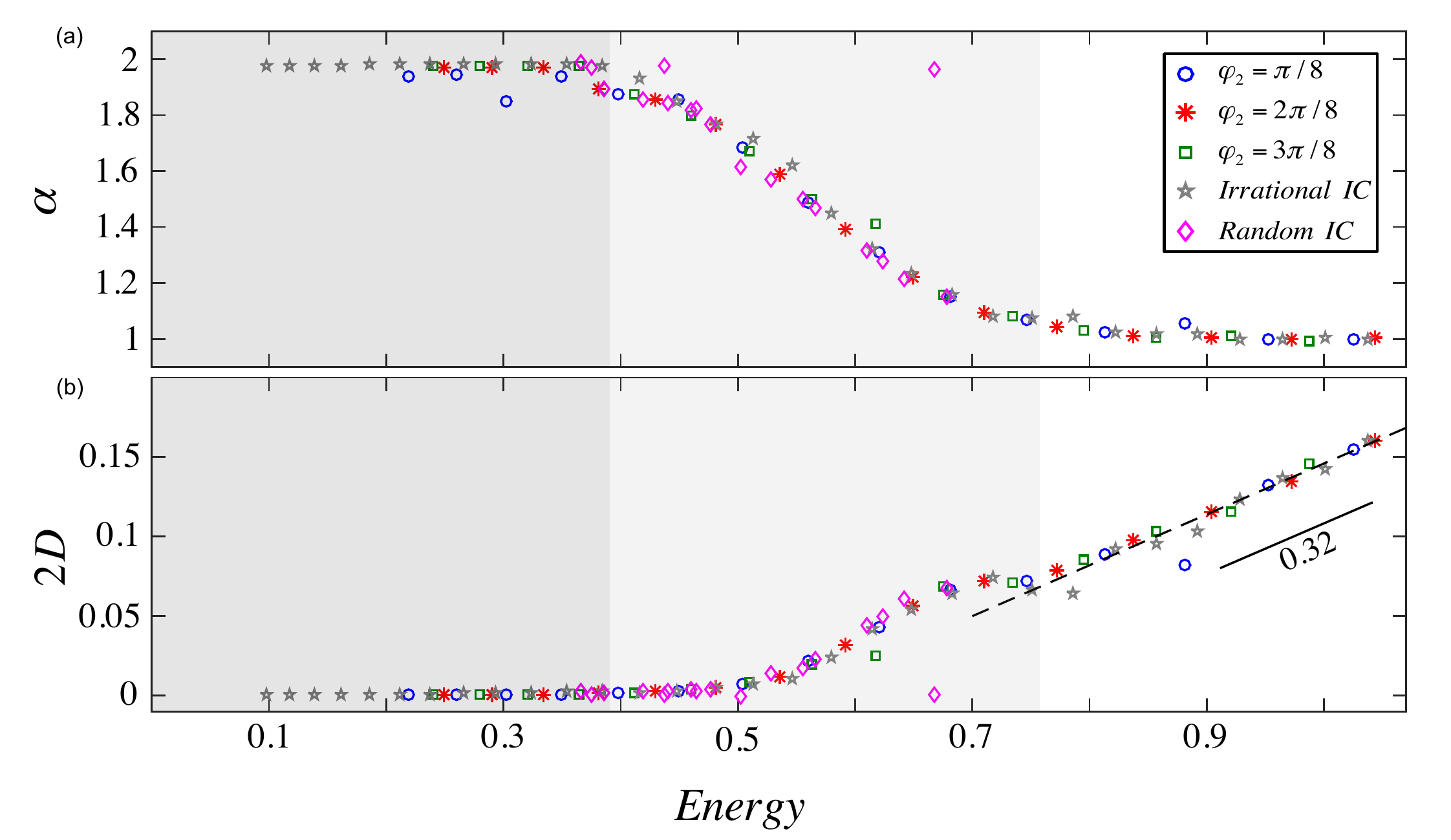}}
\caption{The anomalous diffusion exponent $\alpha$ and the
diffusion coefficient $D$ as a function of the energy for various
types of initial conditions. Error bars are not shown because they
are smaller than the markers. The (blue) circle, (red) asterix and (green)
square markers correspond to initial conditions set by different phase
differences between the degenerate modes. The (gray) pentagram plot
corresponds to irrational initial phase difference. The (magenta) diamond
plot corresponds to random initial conditions. For most initial conditions
the energy suffices for predicting alpha. At around $E=0.35$ trajectories
begin to change from ballistic angular motion to anomalous diffusion
(dark gray and light gray background), and at around $E=0.7$ trajectories begin
to change to regular diffusion (clear background). 
In the regular diffusion regime, where $\alpha=1$, the diffusion
coefficient appears to be linear in the energy with a slope $0.32$.
}\label{fig:alpha&D}
\end{figure*}

A general low energy initial condition will thus evolve according to
\[
w_{i}=A_{i} \cos(\omega_{i} t+\varphi_{i}).
\] 
Unlike the small displacement approximation, in the present case calculating the full motion resulting from these oscillation yields a non vanishing average angular velocity given by \cite{Iwai2005}
\begin{equation}
\bar{\theta}_{1}\left(t\right)=\epsilon^{2}\frac{1}{16}\sqrt{\frac{3}{2}}A_{1}A_{2}\sin(\varphi_{1}-\varphi_{2})\,t .
\label{eq:constVel}
\end{equation}
This ratcheting constant average angular velocity motion with zero angular momentum is a direct outcome of the degeneracy of the spectrum of the problem. If one, for example, considers a system with one of the masses changed this degeneracy is lifted and no cumulative angular motion is observed at long times in the linear regime.
It is thus somewhat surprising that the constant angular velocity persists for non-linear oscillations where the frequencies of the degenerate modes attempt to diverge, and the angular motion displays beating. Indeed, in this non-linear regime the prediction of the averaged angular velocity \eqref{eq:constVel} increasingly deviates from the observed one as the energy is increased and the interaction between the modes becomes more pronounced. These non-linear oscillations are well captured by Birkhoff normal form theory, presented in \cite{Bambusi2014} and applied to the harmonic three body in \cite{KE18}, where the the beating observed in the dynamics of $\theta_{1}$ is identified with periodic energy transfer between the degenerate modes.

For moderate energies the regular nature of the dynamics of the system breaks. This in particular manifests in the ``filling up" of the spectrum observed in Figure \ref{fig:trajectories}. 
For short times the motion is reminiscent of the regular motion observed for lower energy; the system displays non-linear oscillations superimposed on a constant angular velocity motion. However, careful examination reveals irregularities in the constant average angular velocity motion that were not present for lower energy excitations. After a finite time the direction of the rotation abruptly reverses and the system rotates, again with a constant angular velocity, in the opposite direction. The durations of the constant angular velocity bouts are random, and for low enough energies the tail of their distribution obeys a power law: $P(\tau)\propto \tau ^{-\nu-1}$. This is the exact setting of the so-called L{\'{e}}vy-walk model \cite{Geisel1992} used to describe anomalous sub and super-diffusive phenomena in systems ranging from quantum transport \cite{Lee2009}, through turbulence \cite{Shlesinger1987} to biological locomotion \cite{Ariel2015}. For values of $1<\nu<2$ this leads to an average squared mean angular displacement exponent $\left< \theta^{2}\right>\propto t^\alpha$ that satisfies $\alpha=3-\nu$. Near the loss of integrability the obtained values of $\nu$ is close to $1$ and $\alpha$ follows this prediction well (see Figure \ref{fig:Levy}). However, as the energy is increased the bout distribution exponent $\nu$ increases away from $1$ and the fit to a clean power law deteriorates. In this regime the square mean displacement exponent $\alpha$ also deviates from the L{\'{e}}vy-walk predicted value, yet it continues to decrease monotonically until reaching the value $\alpha=1$ characterizing a standard random walk. We note that the energy for which the mean squared displacement exponent reaches the value $\alpha = 1$ is only slightly above the threshold energy for collinear configurations $E_{t}=2/3$, beyond which the entirety of the equienergy phase space becomes accessible. Such collinear configurations, followed by orientation reversal of the spring mass triangle, are rare at $E\gtrsim 2/3$ and occur only a few times during a typical run of $10^7$ typical times. As the energy is increased such orientation reversals become more and more frequent. However, the angle $\theta_1$ is still well defined and strobing its value only whenever the triangle returns close enough to its original unreflected orientation produces
the same temporal evolution. For sufficiently high energy the square mean displacement scales linearly with time in agreement with the predicted exponent of an angular random walk. It is however important to note that the agreement with the statistics of angular random walk is not observed for the higher moments of the angular displacement, see SI.   

We note that for vanishing rest lengths the potential becomes purely harmonic and the system recovers regularity, see SI and \cite{CS93}. We expect the system to approach this limit when provided with extremely high energy and the typical displacement satisfies $L\ll|r_{ij}|$, yet we have not throughly explored this transition.

The harmonic three mass problem shows a remarkable variety of behaviors with the total energy content of the system as the main control parameter. As the system is mixed, islands of regular behavior are expected to be found for every value of the total energy in the system. However, these islands become small very fast as the energy of the system is increased allowing us to study a "typical" behavior for a given value of the energy. This in particular manifests in the data collapse of the fractional diffusion exponent $\alpha$ and of the diffusion coefficient $D$ as a function of the energy for a variety of initial conditions. Occasionally we numerically find the regular islands (an example is observed in Figure \ref{fig:alpha&D} where a random initial condition at the energy of almost $0.7$ leads to a ballistic statistics with $\a=2$), yet these are scarce.

During the L{\'{e}}vy walk regime the system displays random orientation reversals, spending long periods of time in a trajectory very close to the regular quasi-periodic trajectory of constant averaged angular velocity, followed by an abrupt transition to the vicinity of the oppositely rotating quasi-periodic trajectory. 
A somewhat similar scenario was observed in \cite{ZN97,AFM14} studying the kicked rotor Hamiltonian and its discrete map realizations. The phase spaces of these systems display islands of regular motion corresponding to ballistic propagation of the relevant coordinate or its conjugate momentum. These are non-autonomous systems and the islands of regular motion in them may appear for certain values of the relevant parameters rather than persist from the linear theory as in our case. Nonetheless, there are many similarities between the systems. In particular, trajectories outside the regular motion islands tend to stick to them and accordingly follow a path of ballistic propagation for a finite time before leaving their vicinity.  The distribution of the times spent in the vicinity of the islands gives rise to the anomalous statistics of the motion.

It is presently unclear what is the source of the non-trivial distribution of sticking times in our system. One plausible explanation attributes the anomalous statistics to the phase space structure at energies at which critical KAM tori are destroyed by perturbations. As has been observed in several low-dimensional mixed systems \cite{MMP84,ArnoldDynamical1988}, as resonant tori are destroyed they can break up into an infinite hierarchy of smaller islands. Such nested islands create a self-similar sequence of partial boundaries in which an orbit remains trapped for long times before escaping. The fractal structure of the boundaries results in L{\'{e}}vy statistics of trapping times \cite{Geisel1987}.
While the Poincar{\'{e}} sections at different energy regimes further strengthen this interpretation of the phase space structure (see SI), further studies of the system are required in order to establish the relevance of this explanation to the phenomenon observed. 

As the energy grows and the integrable islands shrink, the sticking times shorten. Correspondingly, the MSD exponent $\alpha$ seems to vary continuously from almost 2 near the ballistic regime, to 1 at sufficiently high energies fitting angular random walk statistics. Poincar{\'{e}} sections in the high energy regime show a seemingly random distribution of points. However, much like in the Fermi-Pasta-Ulam system, it is not clear whether or not the system ever reaches a thermal equilibrium in which the energy is truly equipartitioned between the different modes \cite{Gal07}. Further, the deviation of higher order moments of the angular displacement from the expected linear relation of an uncorrelated random walk model (see SI) indicates that the dynamics retain some correlations. 
It is however plausible that the system at sufficiently high energies could be considered as serving as its own thermal bath. In this case it may provide insight for the transition of a many particle autonomous Hamiltonian system from a ``localized" regular state to a thermalized uncorrelated state \cite{NH15,AV15}.

As the evolution of the non-holonomic variable depends on the full history of the dynamics of the independent variables in the system it serves as an exceptionally good proxy for temporal correlations in the trajectory of the independent variables. In the present case theses correlations manifest in the non-trivial statistics of the angular displacement of the three mass triangle, yet other systems may display these correlations through different modes of motion. For example we may consider the harmonic three body system constrained to curved space. The motion of isolated particles in spaces of uniform Riemannian curvature obeys the conservation of linear momentum and such bodies cannot acquire momentum. However, much like the system presented here which allows rotations with vanishing angular momentum, in curved space non-rigid bodies can translate without linear momentum. This type of motion was termed swimming in curved space by Wisdom \cite{Wisdom2003} and later implemented by Avron and Kenneth for the case of a three point mass swimmer with controllable connectors \cite{Avron2006}. If the three-mass harmonic system is to be solved in a uniformly curved geometry we predict that the resulting internal dynamics described here will manifest as real space diffusion, actualizing Robert Brown's initial interpretation of a random walk as belonging to the particle itself. The exact nature of this motion is yet to be explored. 

\newpage

\appendix
\section{Supplementary Information}

\subsection{Methods of Simulation}\label{App:simulation}

The simulations were performed using the semi-implicit Euler method, a
symplectic integration method that preserves the constants of motion
of the Hamiltonian \cite{Hairer2015}. The simulations were run for typical times of
$10^{7}$ with a time step of $dt=0.01$. Initial conditions were
always chosen with zero angular and linear momentum, and the center
of mass of the system at the origin. They were determined by using
the normal modes from the linearization, $w_{i}=A_{i}\cos\left(\omega_{i}t+\varphi_{i}\right)$,
subscribing values for $\left\{ A_{i},\varphi_{i}\right\} _{i=1}^{3}$,
and mapping back to Cartesian coordinates.

Simulating a chaotic system over long periods of time is a subtle
issue \cite{Yao2010}. As opposed to non-chaotic systems, which present convergence
of a numerical solution to the real solution as the resolution of
the simulation is improved, in chaotic systems no such convergence
is obtained. Any change in the numerical method, working precision,
time step, and of course initial conditions, results in radically
diverging trajectories because of the sensitive dependence of chaotic
systems on their exact parameters. Therefore, any numerical method
simulating the behavior of a chaotic system is doomed to failure after
a finite time, and exact long-time simulations of such systems are
practically impossible. The natural question raised is, which physical
quantities, if any, can nevertheless be calculated from numerical
trajectories of a chaotic system?

In performing long-time simulations, we discovered that non-physical
changes in the numerics, e.g. changing the time-step or the integration
method, result in the same qualitative types of trajectories (despite
drastic quantitative differences), see Figure \ref{fig:MSDcalc}. In the intermediate
energy regime, where the motion is described by constant angular velocity
bouts of motion that switch direction after random times, we find
that although the exact point at which a new bout will commence depends
strongly on the non-physical parameters, over long times the PDF of
the bout lengths stays the same. Furthermore, the power spectrum looks
the same, and the velocities that the bouts bounce between stay approximately
the same. In the high energy regime, where the motion
resembles an angular random walk, we see that trajectories differing
in non-physical parameters look similar, separating exponentially
in a manner resembling infinitesimally close trajectories. Most importantly,
the MSD of the motion is invariant to these changes.

Thus it seems that by using the symplectic integrator, trajectories
stay in the same ensemble of motion despite non-physical changes.
Subsequently, we hypothesize that statistical averages performed on our
system represent the true dynamics, despite each individual trajectory
diverging from the true trajectory. This of course does not constitute
a proof, but the robustness of the MSD calculation and the smoothness
in the obtained values of the diffusion coefficients and exponents
do render this approach very plausible. 

\subsection{Cartesian Linearization}\label{App:lin}

The full symmetric system, presented in equation \eqref{eq:hamiltonian}, has an equilibrium
position when the triangle is equilateral and its sides, the distances
between the masses, equal the rest length, i.e. when $r_{ij}=L$ for
all $\left\langle i,j\right\rangle $. In the Cartesian coordinates,
there are countless configurations $\left\{ \vec{r}_{i}\right\} _{i=1}^{3}$
that correspond to this equilibrium, differing by rotations, translations
and reflections. In order to linearize about the equilibrium, one
of these configurations must be chosen. The solution to this exercise also
appears in \cite{Kotkin1971}.

We shall linearize around the configuration positioned such that the
center of mass is at the origin and $\vec{r}_{23}$ is parallel to
the $x$-axis: 
\[
\vec{r}_{i}=\vec{r_{i}^{0}}+\epsilon\vec{\delta r_{i}},
\]
\[
\vec{r_{1}^{0}}=\left(0,\frac{\sqrt{3}}{3}L\right)\,;\,\vec{r_{2}^{0}}=\left(-\frac{L}{2},-\frac{\sqrt{3}}{6}L\right)\,;\,\vec{r_{3}^{0}}=\left(\frac{L}{2},-\frac{\sqrt{3}}{6}L\right).
\]

The linearized equations of motion are given by:
\[
\left(\begin{matrix}\ddot{\delta x}_{1}\\
\ddot{\delta x}_{2}\\
\ddot{\delta x}_{3}\\
\ddot{\delta y}_{1}\\
\ddot{\delta y}_{2}\\
\ddot{\delta y}_{3}
\end{matrix}\right)=\frac{k}{4m}\left(\begin{matrix}-2 & 1 & 1 & 0 & \sqrt{3} & -\sqrt{3}\\
1 & -5 & 4 & \sqrt{3} & -\sqrt{3} & 0\\
1 & 4 & -5 & -\sqrt{3} & 0 & \sqrt{3}\\
0 & \sqrt{3} & -\sqrt{3} & -6 & 3 & 3\\
\sqrt{3} & -\sqrt{3} & 0 & 3 & -3 & 0\\
-\sqrt{3} & 0 & \sqrt{3} & 3 & 0 & -3
\end{matrix}\right)\left(\begin{matrix}\delta x_{1}\\
\delta x_{2}\\
\delta x_{3}\\
\delta y_{1}\\
\delta y_{2}\\
\delta y_{3}
\end{matrix}\right)\,.
\]

The solutions are $\vec{r}_{i}=\vec{r_{i}^{0}}+\epsilon\vec{\delta r_{i}}$,
where $\vec{\delta r}=B^{-1}\vec{u}$ and:
\[
\begin{aligned}
&B&=&\left(\begin{array}{cccccc}
0 & \sqrt{3} & -\sqrt{3} & -2 & 1 & 1\\
\frac{1}{\sqrt{3}} & -\frac{2}{\sqrt{3}} & \frac{1}{\sqrt{3}} & -1 & 0 & 1\\
-\frac{1}{\sqrt{3}} & -\frac{1}{\sqrt{3}} & \frac{2}{\sqrt{3}} & -1 & 1 & 0\\
-\frac{\sqrt{3}}{2} & 0 & 0 & \frac{1}{2} & 0 & 1\\
\frac{\sqrt{3}}{2} & 0 & 0 & \frac{1}{2} & 1 & 0\\
1 & 1 & 1 & 0 & 0 & 0
\end{array}\right)\,; \\
\,&\left(\begin{matrix}u_{1}\\
u_{2}\\
u_{3}\\
u_{4}\\
u_{5}\\
u_{6}
\end{matrix}\right)&=&\left(\begin{matrix}A_{1}\cos\left(\sqrt{3}\frac{t}{\tau}+\phi_{1}\right)\\
A_{2}\cos\left(\sqrt{\frac{3}{2}}\frac{t}{\tau}+\phi_{2}\right)\\
A_{3}\cos\left(\sqrt{\frac{3}{2}}\frac{t}{\tau}+\phi_{3}\right)\\
a_{4}t+b_{4}\\
a_{5}t+b_{5}\\
a_{6}t+b_{6}
\end{matrix}\right).
\end{aligned}
\]

The directions of each of the oscillatory normal modes $u_{1}$, $u_{2}$
and $u_{3}$ are marked in Figure \ref{fig:spring-mass}(b, c, d) respectively. The modes
which are linear with $t$ -- $u_{4}$, $u_{5}$ and $u_{6}$ -- vanish
in the center of mass, zero angular momentum frame. We can see that
the normal modes correspond to the known molecular vibrations of water
- the $\sqrt{3}$ frequency of $u_{1}$ corresponds to the symmetric
stretch, while the $\sqrt{3/2}$ frequency of $u_{2}$ and $u_{3}$
corresponds to the asymmetric stretch and the symmetric bend.

\subsection{Reduction to Shape Space}\label{App:shape}

In this section we formulate the system as a deforming triangle placed
in the plane instead of as three masses moving independently, where
we differentiate between reflection-related configurations despite
the symmetry in parameters. To this end we employ a change of variables
from the initial six Cartesian position variables to three shape-space
variables that determine the shape of the triangle, and three configuration-space
variables that describe the position and orientation of the triangle
in the plane. As we will soon show, the dynamics of the three configuration
space variables are fully determined by the initial conditions, the
three shape space variables and their dynamics. This allows a reduction
of the system to shape space, the subspace of shape variables, resulting
in a reduced 3-degree-of-freedom Lagrangian whose motion fully determines
the full motion of the system. This process is described thoroughly
in \cite{Montgomery2015,Littlejohn1997,Iwai1987}. We describe below
the details most relevant to the present case. We note that the general
setting of this approach can be formulated as a type of gauge theory
for deformable objects \cite{Littlejohn1997}.

We start with equations of motion derived from the Lagrangian depending
on the full space of configurations of the system. A specific configuration
of the system at some time $t$ is given by the positions of the three
masses, a vector set $\left(\vec{r}_{1}\left(t\right),\vec{r}_{2}\left(t\right),\vec{r}_{3}\left(t\right)\right)$,
$\vec{r}_{i}\left(t\right)\in\mathbb{R}^{2}$. We denote this six-dimensional
configuration space as $Q:=\mathbb{R}^{2}\times\mathbb{R}^{2}\times\mathbb{R}^{2}$.
Treating the three-mass system as a deforming triangle, we would like
to separate $Q$ into two subspaces: shape space, a 3D subspace defining
the shape of the triangle denoted by $S$ , and placement space,
defining the placement and orientation of the triangle in the plane.
Thus a certain configuration of the system $\left(\vec{r}_{1},\vec{r}_{2},\vec{r}_{3}\right)$
can be written as $\left(s,\vec{R}_{CM},\theta\right)\in S\times\mathbb{R}^{2}\times[0,2\pi)$,
where $s\in S$ signifies a certain triangle , $\vec{R}_{CM}$ denotes
the position of the center of mass of the triangle and $\theta$ marks
its orientation around $\vec{R}_{CM}$ . We note that in order for
this transformation to be well defined we need the shape variables
$s$ to differentiate between clockwise and counterclockwise configurations
of the masses. This means that two triangles placed in the plane have
the same shape if there is a rigid motion, i.e. a composition of translation
and rotation without reflection, which relates them  \cite{Montgomery2015}.

\begin{figure}[ht]
\centerline{\includegraphics[width=.35\textwidth]{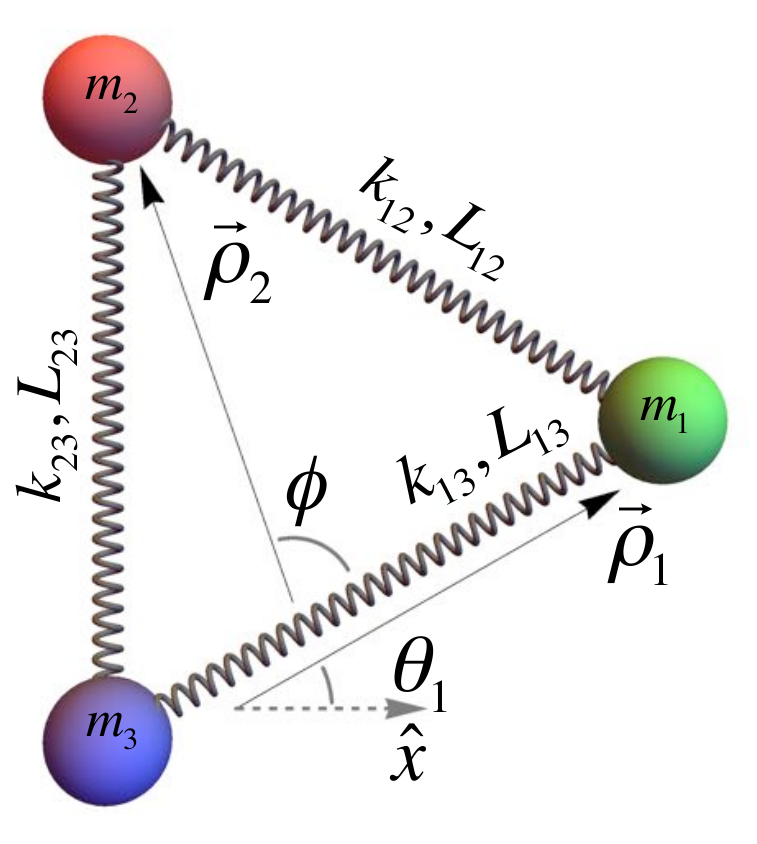}}
\caption{The harmonic three body system in Jacobi coordinates. $\rho_{1}$ and $\rho_{2}$ are the two relative Jacobi vectors. $\theta_1$ is the angle between $\rho1$ and the $x$-axis, and is used to determine the orientation of the triangle in the plane. $\phi$ is the angle between the two Jacobi vectors, and together with the lengths $|\rho_{1}|$ and $|\rho_{2}|$ is used determine the shape of the triangle.}
\label{fig:jacobi}
\end{figure}

The separation is achieved in two steps. First, the center of mass
variables are eliminated by formulating the problem in the scaled
Jacobi coordinates \cite{Montgomery2015,Littlejohn1997}, consisting
of two relative coordinates $\vec{\rho}_{1},\vec{\rho}_{2}$ and the
center-of-mass coordinate $\vec{R}_{CM}$ (see Figure \ref{fig:jacobi}):

\[
\begin{aligned}\vec{\rho}_{1} =\left(\frac{1}{m_{1}}+\frac{1}{m_{2}}\right)^{-1/2}\left(\vec{r}_{1}-\vec{r}_{2}\right),\\
\vec{\rho}_{2} =\left(\frac{1}{m_{3}}+\frac{1}{m_{1}+m_{2}}\right)^{-1/2}\left(\vec{r}_{3}-\frac{m_{1}\vec{r}_{1}+m_{3}\vec{r}_{2}}{m_{1}+m_{3}}\right),\\
\vec{R}_{CM} =\frac{\sum_{i}m_{i}\vec{r}_{i}}{m_{1}+m_{2}+m_{3}}.
\end{aligned}
\]

In these coordinates, the potential energy depends only on the relative
coordinates $\vec{\rho}_{1}$, $\vec{\rho}_{2}$, while $\vec{R}_{CM}$
decouples from the rest of the Lagrangian. We are thus allowed, without
the loss of generality, to analyze the system in the center of mass
frame. This constitutes a reduction of the configuration space to
four dimensions, with a configuration described by $\left(\vec{\rho}_{1},\vec{\rho}_{2}\right)\in\mathbb{R}^{2}\times\mathbb{R}^{2}$.
Furthermore, $\vec{R}_{CM}$ is a cyclic coordinate of the Lagrangian
and determined completely by the linear momentum of the system $\vec{P}$
and its initial conditions: $\vec{R}_{CM}\left(t\right)=\left(m_{1}+m_{2}+m_{3}\right)^{-1}\vec{P}t+\vec{R}_{CM}\left(0\right)$.

Next, following the work of \cite{Iwai1987} we carry out a final change
of variables to eliminate the orientation variable:
\[
\begin{aligned}w_{1} & =\frac{1}{2}\left(\rho_{1}^{2}-\rho_{2}^{2}\right),\\
w_{2} & =\vec{\rho}_{1}\cdot\vec{\rho}_{2}=\rho_{1}\rho_{2}\cos\phi,\\
w_{3} & =-\vec{\rho}_{1}\wedge\vec{\rho}_{2}=-\rho_{1}\rho_{2}\sin\phi,
\end{aligned}
\]
where $\vec{\rho}_{i}=\left(\rho_{i}\cos\theta_{i},\rho_{i}\sin\theta_{i}\right)$
and $\phi=\theta_{2}-\theta_{1}$ . The above components can be written
in spherical coordinates: 

\[
\vec{w}=\left(w_{1},w_{2},w_{3}\right)=\left(w\sin\chi\cos\psi,w\sin\chi\sin\psi,w\cos\chi\right).
\]

\begin{figure}[ht]
\centerline{\includegraphics[width=.5\textwidth]{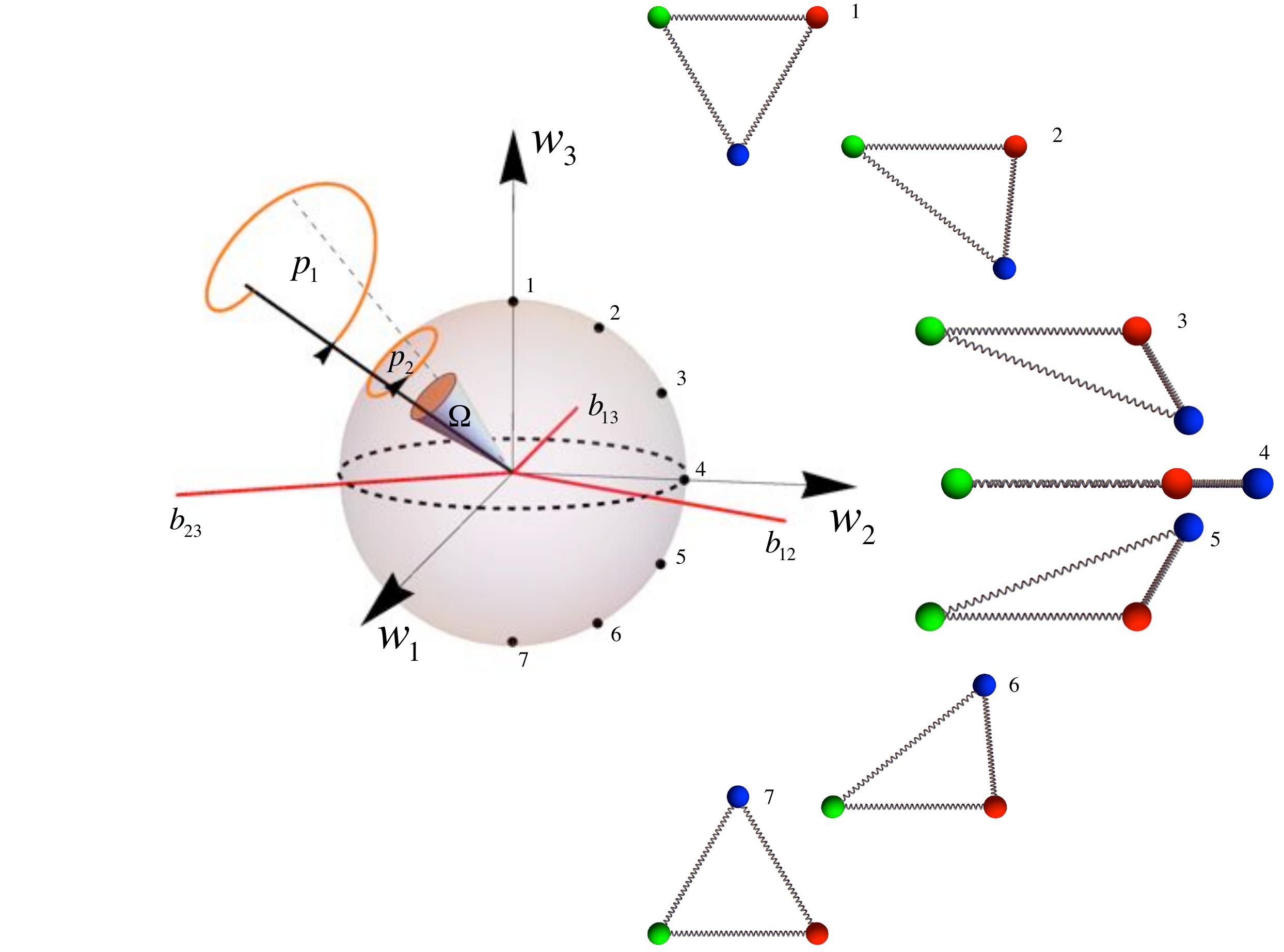}}
\caption{Shape space. Every point in this three-dimensional space corresponds to a different
triangle shape. Similar triangles sit on the same ray projecting from
the origin, so the shape sphere (marked in the figure) contains the
full set of possible triangles up to similarity, excluding the three-point
collision. A few examples of points and their corresponding triangles
are marked in the figure (1-7). The red rays, marked $\left\{ \vec{b}_{ij}\right\} _{\left\langle i,j\right\rangle }$,
are the collision rays, which describe two-point collision points
between $m_{i}$ and $m_{j}$. $p_{1}$ describes some trajectory
of the system for which the triangles at the beginning and at the
end are similar, sitting on the same ray extending from the origin.
$p_{2}$ describes the projection of the trajectory to the shape sphere.
As explained in the text, the solid angle enclosed by $p_{2}$ is
proportional to the rotation the triangle acquires at the end of this
trajectory, assuming it has zero angular momentum throughout the motion.}
\label{fig:shapeSpace}
\end{figure}

These coordinates constitute our three-dimensional shape space $S$.
Each $w_{i}$ can attain the full range $w_{i}\in\mathbb{R}$, so
$S\overset{\sim}{=}\mathbb{R}^{3}$. Every point in shape space $\vec{w}\in S$
corresponds to a unique triangle shape (Figure \ref{fig:shapeSpace}). Also, since the potential
energy depends only on the shape of the triangle, it can be fully
expressed using the shape space coordinates $\vec{w}$.

The variables $w_{1}$ and $w_{2}$ don't have a straightforward interpretation
in terms of the triangle's shape, however $w_{3}$ does have an intuitive
interpretation: it is proportional to the signed area of the triangle,
positive if the masses $m_{1}$, $m_{2}$ and $m_{3}$ are ordered
clockwise and negative if they are ordered counter-clockwise. Thus
two triangles $\left(w_{1},w_{2},w_{3}\right)$ and $\left(w_{1},w_{2},-w_{3}\right)$
are related by reflection. The plane $w_{3}=0$ corresponds to collinear
configurations, for which the area is zero, and the origin $\left(0,0,0\right)$
is the three-point collision point. On the collinear plane, starting
from $\left(0,0,0\right)$ three rays extend, which describe two-point
collision points. We define the corresponding normalized collision
vectors $\left\{ \vec{b}_{ij}\right\} _{\left\langle i,j\right\rangle }$,
with values depending on the masses of the system (equation \eqref{eq:collisionRay}). Thus
a point $\vec{w}=\alpha\vec{b}_{ij}$ for $\alpha\in\mathbb{R}^{+}$
describes a triangle in which $m_{i}$ and $m_{j}$ collide. 

The size of a shape vector $w:=\left|\vec{w}\right|$ determines the
scaling of the triangle. Similar triangles will share $\chi$ and
$\psi$ and differ in $w$ \cite{Montgomery2015}. Therefore, each
centered sphere $S_{a}=\left\{ \vec{w}:\,w=a>0\right\} $ contains
the full set of possible triangles up to similarity, excluding the
three-point collision. In other words, each triangle $\vec{w}\in S\backslash\left\{ 0,0,0\right\} $
is similar to one specific triangle $\vec{w}_{a}\in S_{a}$, and furthermore
they both sit on the same ray extending from the origin . Specifically,
the unit sphere $S_{1}=\left\{ \vec{w}\,:\,w=1\right\} $ can be mapped
to the congruence class of all similar triangles. This sphere is termed
the \textquotedblleft shape sphere\textquotedblright{} \cite{Montgomery2015}. 

At this point, we have achieved the desired separation of the full
configuration space: a general configuration $\left\{ \vec{r}_{i}\right\} _{i=1}^{3}$
is mapped to a unique set $\left(\vec{w},\vec{R}_{CM},\theta_{1}\right)$.
However, there are many different possible ways we could assign an
orientation to the triangle other than $\theta_{1}$ . For example,
we could define the orientation of the triangle using the angle of
one of the masses with relation to the $\hat{x}$-axis, the orientation
of one of the triangle sides with respect to the $\hat{y}$-axis,
or any other choice. Furthermore, we could define an orientation depending
on the shape of the triangle. Any such choice is arbitrary, and can
be viewed as a type of gauge choice: since a shape holds no meaning
of orientation, a mapping must be made between shapes $\vec{w}\in S$
and configurations $\left\{ \vec{r}_{i}\right\} _{i=1}^{3}\in Q$
that place a triangle at a specific orientation and position in the
plane, and there is freedom in this mapping to decide how the orientation
and position are defined. It is clear by this formulation that the
gauge choice is a redundancy of the reduction, and has no real physical
meaning. Therefore we demand from any physical quantity that we would
like to calculate to be gauge invariant \cite{Littlejohn1997}.

At this point we can return to the Lagrangian, changing variables
again from $\vec{\rho}_{1},\vec{\rho}_{2}$ to $\vec{w},\theta_{1}$.
In doing so we note that the Lagrangian depends on the orientation
$\theta_{1}$ only through the angular velocity $\dot{\theta}_{1}$.
This is a manifestation of the angular momentum conservation built
in to the problem: the system is invariant to rotations. The angular
momentum itself is given by:

\[
L=\frac{\partial\mathcal{L}}{\partial\dot{\theta}_{1}}=\frac{w_{3}\dot{w}_{2}-w_{2}\dot{w}_{3}}{w+w_{1}}+2w\dot{\theta}_{1}.
\]

By setting the angular momentum to zero, we obtain an equation relating
the angular velocity $\dot{\theta}_{1}$ with the shape coordinates:
\[
\dot{\theta}_{1}=\frac{\dot{w}_{3}w_{2}-\dot{w}_{2}w_{3}}{2w\left(w+w_{1}\right)}=-\frac{1}{2}\frac{\dot{\chi}\sin\psi+\dot{\psi}\sin\chi\cos\chi\cos\psi}{1+\sin\chi\cos\psi}.
\]

In differential form:

\[
\begin{aligned}d\theta_{1} & =F_{2}\left(w_{1},w_{2},w_{3}\right)dw_{2}+F_{3}\left(w_{1},w_{2},w_{3}\right)dw_{3}\\
 & =G_{\chi}\left(\chi,\psi\right)d\chi+G_{\psi}\left(\chi,\psi\right)d\psi,
\end{aligned}
\]
where:

\[
\begin{aligned}
&F_{2}\left(w_{1},w_{2},w_{3}\right)=-\frac{w_{3}}{2w\left(w+w_{1}\right)}\,,\\
&F_{3}\left(w_{1},w_{2},w_{3}\right)=\frac{w_{2}}{2w\left(w+w_{1}\right)},\\
&G_{\chi}\left(\chi,\psi\right)=-\frac{\sin\psi}{2\left(1+\sin\chi\cos\psi\right)}\,,\\
&G_{\psi}\left(\chi,\psi\right)=-\frac{\sin\chi\cos\chi\cos\psi}{2\left(1+\sin\chi\cos\psi\right)}.
\end{aligned}
\]

This equation is the connection form \cite{Montgomery1993}, and contains
the information on how to connect two infinitesimally close shapes
such that the motion between them will have zero angular momentum.

We note that the rotational connection does not depend on the scaling
of the triangles, encapsulated by $w$, and ``lives'' on the shape
sphere. This is no surprise, as a rescaling of the triangle carries
no rotational charge.

Finally, by determining zero angular momentum for the entire motion
we can plug the equation for $\dot{\theta}_{1}$ \eqref{eq:theta1} back into the Lagrangian
to obtain the fully reduced Lagrangian, defined on shape space:

\[
\mathcal{L}_{red}=\frac{\dot{w}_{i}^{2}}{4w}-\frac{1}{2}k_{ij}\left(r_{ij}\left(\vec{w}\right)-L_{ij}\right)^{2},
\]
where:

\[
\begin{aligned}
r_{ij}&=\sqrt{\frac{m_{i}+m_{j}}{m_{i}m_{j}}\left(w-\vec{w}\cdot\vec{b_{ij}}\right)}\,\,,\,\,\vec{b_{12}}=\left(\begin{matrix}-1\\
0\\
0
\end{matrix}\right)\,\,,\\
\vec{b_{13}}&=\left(\begin{matrix}\frac{Mm_{1}-m_{2}m_{3}}{\left(m_{1}+m_{2}\right)\left(m_{1}+m_{3}\right)}\\
\frac{2\sqrt{Mm_{1}m_{2}m_{3}}}{\left(m_{1}+m_{2}\right)\left(m_{1}+m_{3}\right)}\\
0
\end{matrix}\right)\,\, ,\,\,\vec{b_{23}}=\left(\begin{matrix}\frac{Mm_{2}-m_{1}m_{3}}{\left(m_{1}+m_{2}\right)\left(m_{2}+m_{3}\right)}\\
-\frac{2\sqrt{Mm_{1}m_{2}m_{3}}}{\left(m_{1}+m_{2}\right)\left(m_{2}+m_{3}\right)}\\
0
\end{matrix}\right).
\end{aligned}
\label{eq:collisionRay}
\]

The sum over $i$, $\left\langle i,j\right\rangle $ is implicit,
and we define $M:=m_{1}+m_{2}+m_{3}$.

Specifically for the case of zero angular momentum, this process is
a simple case of the Routh method for eliminating cyclic coordinates
\cite{ArnoldDynamical1988, Marsden2000} and describes the same motion
as the full system; in \cite{ArnoldDynamical1988} Arnold presents
a simple proof to the statement that a function $\left\{ \vec{r}_{i}\left(t\right)\right\} _{i=1}^{3}$
is a motion of the full system with zero angular momentum if and only
if its projection $\vec{w}\left(t\right)$ is a motion of the reduced
system. Given such a solution of the reduced system $\vec{w}\left(t\right)$,
to recover the full dynamics of the triangle we are required to integrate
$\dot{\theta}_{1}$ and determine the temporal evolution of $\theta_{1}$:

\[
\begin{aligned}\theta_{1}\left(t\right) & =\int_{\vec{w}\left(t\right)}F_{2}\left(w_{1},w_{2},w_{3}\right)dw_{2}+F_{3}\left(w_{1},w_{2},w_{3}\right)dw_{3}\\
 & =\int_{\vec{w}\left(t\right)}G_{\chi}\left(\chi,\psi\right)d\chi+G_{\psi}\left(\chi,\psi\right)d\psi
\end{aligned}
\]

Since these are not conservative vector fields, this integral is path-dependent.
It is this path-dependance which allows DIR . We note that at this
point $\theta_{1}\left(t\right)$ is not a gauge-invariant quantity
\cite{Littlejohn1997}, however the full configuration obtained $\left\{ \vec{r}_{i}\left(\vec{w},\theta_{1}\right)\right\} _{i=1}^{3}$
is gauge invariant.

For deformation sequences whose projection on the shape sphere is
a closed path we can use Green's theorem. This reduces
the change in $\theta_{1}$ to the solid angle the path encloses on
the shape sphere, elucidating its nature as a rotationally symmetric
geometric phase of shape space:

\[
\begin{aligned}\Delta\theta_{1}\left(T\right) & =\iint\left(\frac{\partial G_{\chi}}{\partial\psi}-\frac{\partial G_{\psi}}{\partial\chi}\right)d\chi d\psi\\
 & =-\frac{1}{2}\iint\sin\chi d\chi d\psi=-\frac{1}{2}\iint d\Omega,
\end{aligned}
\]

where the sign of $\Delta\theta_{1}$ is determined by the orientation
of the trajectory forming the boundary of the solid angle. We note
that a closed path on the shape sphere does not necessarily correspond
to a closed trajectory in shape space. It only implies that the triangles
at the beginning and end points are similar, mapping to the same point
on the shape sphere. Similar triangles' orientations compare in a
gauge-invariant way, and indeed the closed-path rotation $\Delta\theta_{1}$
is gauge invariant \cite{Littlejohn1997}. Thus, studying the dynamics
of $\theta_{1}\left(t\right)$ is equivalent to studying the areas
of closed sections on the shape sphere.

This entire process can be put in the broader frame of a classical
gauge theory \cite{Montgomery1993}, in which the shape space plays
the role of the physical space, the symmetry group is $SO\left(2\right)$
and the gauge freedom refers to the freedom in the definition of the
orientation of a shape, which is the freedom in the mapping $S\rightarrow Q$.
The gauge potential is given by the connection field, $\mathbf{A}=\left(G_{\chi},G_{\psi}\right)$,
and the curvature form is given by their two-dimensional curl, $B=\frac{\partial G_{\chi}}{\partial\psi}-\frac{\partial G_{\psi}}{\partial\chi}$
\cite{Littlejohn1997}. Accordingly, the question of whether or not
DIR can occur reduces to the question of the holonomy of the connection;
non-zero curvature is the cause for DIR. In fact, the reduction of
the center-of-mass coordinates can be put in a similar formulation,
the only difference being that the corresponding curvature is zero.

For the symmetric system presented in section 2.2, with which we concern
ourselves here, the relevant equations are:

\[
\mathcal{L}_{red}=\frac{\dot{w}_{i}^{2}}{4w}-\frac{k}{2}\left(r_{ij}\left(\vec{w}\right)-L\right)^{2},
\]
where:
\[
\begin{aligned}
r_{ij}=&\sqrt{\frac{2}{m}\left(w-\vec{w}\cdot\vec{b_{ij}}\right)}\,\,,\,\,\vec{b_{12}}=\left(\begin{matrix}-1\\
0\\
0
\end{matrix}\right)\,\,,\\
\,\,\vec{b_{13}}=&\left(\begin{matrix}1/2\\
\sqrt{3}/2\\
0
\end{matrix}\right)\,\,,\,\,\vec{b_{23}}=\left(\begin{matrix}1/2\\
-\sqrt{3}/2\\
0
\end{matrix}\right).
\end{aligned}
\]

For our further analysis we reformulate the reduced problem as a Hamiltonian
system:

\[
\mathcal{H}_{red}=w\left(p_{1}^{2}+p_{2}^{2}+p_{3}^{2}\right)+\frac{k}{2}\sum_{\left\langle i.j\right\rangle }\left(r_{ij}\left(\vec{w}\right)-L\right)^{2},
\]
where the conjugate momenta $\left\{ p_{i}\right\} _{i=1}^{3}$ are
given by the usual form:

\[
p_{i}=\frac{\partial\mathcal{L}_{red}}{\partial\dot{w}_{i}}=\frac{\dot{w}_{i}}{2w}.
\]

\subsection{Calculating the MSD}\label{App:MSD}

\paragraph{Calculation scheme}

Quantitive analysis of trajectories is performed by calculating the angular
mean-squared displacement (MSD) . In order to obtain enough statistics,
every initial condition for which we would like to calculate the MSD, determined by a unique set $\left\{ A_{i},\varphi_{i}\right\} _{i=1}^{3}$, is simulated ten times, each time with a different infinitesimally
small perturbation to the initial condition of the order of the roundoff
error $\,\sim10^{-8}$. Every simulation is run for a long time $T\sim10^{7}\,$
with a constant time step $dt=0.01$ , where the time units are set
according to the parameters of the simulation: $k=1,L=2,m=1$ and
the corresponding time scale is $\tau=\sqrt{m/k}=1$ . Thus, ten different
angular trajectories $\theta^{i}\left(t\right)\,,\,i=1,...,10$ are
obtained corresponding to the same initial conditions. Then, $\theta^{i}\left(t\right)$
is cut up into pieces of length $L\sim10^{4}$ , and each piece is
treated as a different trajectory in the same ensemble, marking $\theta_{n}^{i}\left(t\right)=\theta^{i}\left(\left(n-1\right)L+t\right)$, with $t\in\left[0,L-1\right]$. Thus we obtain $N=10*T/L\sim10^{4}$
different trajectories. Finally we calculate the MSD and compare it
to a power law:
\[
\sigma^{2}\left(t\right)=\frac{1}{N}\sum_{i=1}^{10}\sum_{n=1}^{T/L}\left(\theta_{n}^{i}\left(t\right)-\theta_{n}^{i}\left(0\right)\right)^{2}=2Dt^{\alpha}.
\]

The justification for this process comes from the simulation's inescapable
error accumulation. At every step the system accumulates a numerical
error deriving from roundoff errors and the time-step error. After
a certain time the system deviates from its original trajectory while
maintaining the same energy and quantitative behavior, thanks to the
use of the symplectic integrator. Therefore every point on the trajectory
can be seen as an initial condition which would produce a perturbed
trajectory in the same ensemble as the original one, where we define
an ensemble as a set of points in an equi-energy hypersurface of the
phase space for which the corresponding trajectories share the same
$\alpha$. 

The correlation time (length of a trajectory segment), taken as $d=5000$
in this analysis, is quite arbitrary, but the resulting MSD is robust
to different correlation times as long as they are long enough to
allow for large $t$'s, and small enough to allow for enough statistics.
Quantitatively we must make sure that $1\ll d\ll T$.

Trajectories for which the velocity autocorrelations decay fast enough,
like the regular random walk process, satisfy $\alpha=1$. This corresponds
to regular (rotationally) diffusive behavior of the ensemble. Constant
ballistic motion, where $\theta\left(t\right)=\theta_{0}+vt$ , produces
$\alpha=2$ and $2D=v^{2}$. Super-diffusive behavior is defined by
motion for which $1<\alpha<2$ .

\begin{figure}[ht]
\centerline{\includegraphics[width=.45\textwidth]{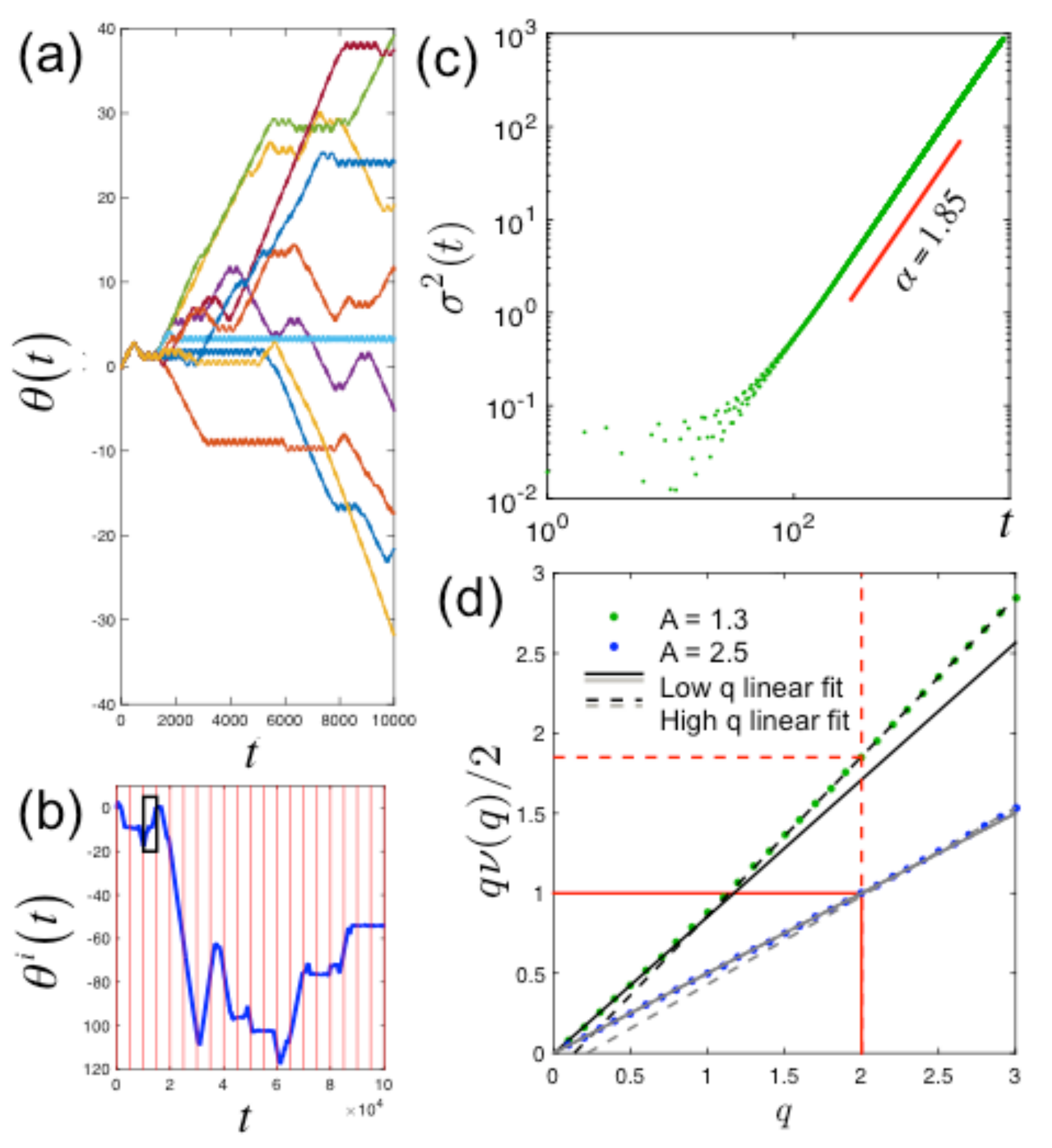}}
\caption{Calculating the MSD and the fractional moments. The calculation is done on 10 trajectories with the initial
conditions : $A_{1}=A_{2}=A_{3}=1.3$ , $\varphi_{1}=\varphi_{3}=0$,
$\varphi_{2}=\pi/8$, each with initial conditions perturbed by an
infinitesimal amount (see first paragraph of this appendix). A part
of the corresponding trajectories $\theta^{i}\left(t\right)$, $i=1,...,10$,
is presented in (a). One of them is shown in (b). Marking every $5000t$,
we cut up the trajectory into 2000 small trajectories (b). Then,
we calculate the MSD of the trajectories according to $\sigma^{2}\left(t\right)=\frac{1}{N}\sum_{n=1}^{N}\left(\theta_{n}\left(t\right)-\theta_{n}\left(0\right)\right)^{2}$,
to obtain (c). The log-log plot shows the power-law is an excellent
fit. We obtain $\alpha=1.85$ from this fit. We see that as expected,
the linear fit is good when $t$ is big enough - for small values
of $t$ the details of the short-term behavior of the trajectory produce
oscillations, which straighten out as $t$ grows. The fractional moments
exhibit the expected bilinear behavior characteristic of L{\'{e}}vy walks
(d).}
\label{fig:MSDcalc}
\end{figure}

An example of the calculation is shown in Figure \ref{fig:MSDcalc}.

\paragraph{Mapping the phase space}

We would like to map the entire phase space using the MSD exponent,
so that every point in phase space can be taken as an initial condition,
for which there exists an ensemble of trajectories as described above.
Therefore, each phase-space point has an anomalous diffusion exponent
$\alpha$ related to it, which marks its regime - $\alpha=2$ for
the periodic regime, $\alpha=1$ for the fully chaotic regime and
$1<\alpha<2$ for the intermediate regime.

We note that a priori there is nothing to stop $\alpha$ from taking
different values than this, but after countless simulations we can
safely say that it would be very surprising to find a regime in which
$\alpha<1$.

However, it is pretty much impossible to map the entire phase space
in this way given its high dimensionality - 6 dimensional phase space
- it's too computationally expensive, and even if it could be done
there isn't a good way to visualize it and obtain relevant information
from it. Therefore, we must ``hand-pick'' specific initial conditions
for which to calculate $\alpha$. Perhaps surprisingly, we find that
the energy is the main parameter controlling the value of $\alpha$,
as can be seen in Figure \ref{fig:alpha&D}. The trajectories for which $\alpha$ deviate
greatly from other values obtained with the same energy are within periodic
islands in the phase space.

\paragraph{Fractional and higher moments}

Fractional moments, used in \cite{Zaburdaev2015} to characterize
the stochastic transport phenomena, can be calculated in a similar
manner. They are a generalization of the MSD, defined as:
\[
S_{q}\left(t\right):=\left\langle \left|\theta\left(t\right)-\theta\left(t\right)\right|^{q}\right\rangle \backsimeq M_{q}\cdot t^{q\nu\left(q\right)/2}.
\]

For normal diffusion, $\,\nu\left(q\right)=1$, whereas if $\,\nu\left(q\right)$
is not constant each of the moments holds new information about the
dynamics \cite{Bouchbinder2006}. This kind of diffusion is referred
to as strongly anomalous. The bilinear behavior observed in Figure \ref{fig:MSDcalc}
is typical for L{\'{e}}vy-walk motion \cite{Zaburdaev2015}. However, even
in the supposedly regular diffusion regime, where $\alpha=1$, we
can still observe a strong bilinear behavior at higher moments, signifying
that the system does not perform a clean random walk but in fact still
retains some non-trivial correlations.

\subsection{Calculating the PDF}\label{App:PDF}

In the intermediate energy regime, the calculated trajectories are
similar to the L{\'{e}}vy-walk model presented in \cite{Zaburdaev2015, Geisel1992}; they're built
from bouts of mostly constant average angular velocity made up from
small oscillations. In order to check the fit to the model, we are
required to calculate the probability density function of the bout
lengths. To this end, we needed to create an algorithm that identifies
the turning-points of a trajectory, a difficult task because of the
small oscillations making up the bouts, and because the model does
not match the observations perfectly.

\begin{figure*}[ht]
\centerline{\includegraphics[width=.75\textwidth]{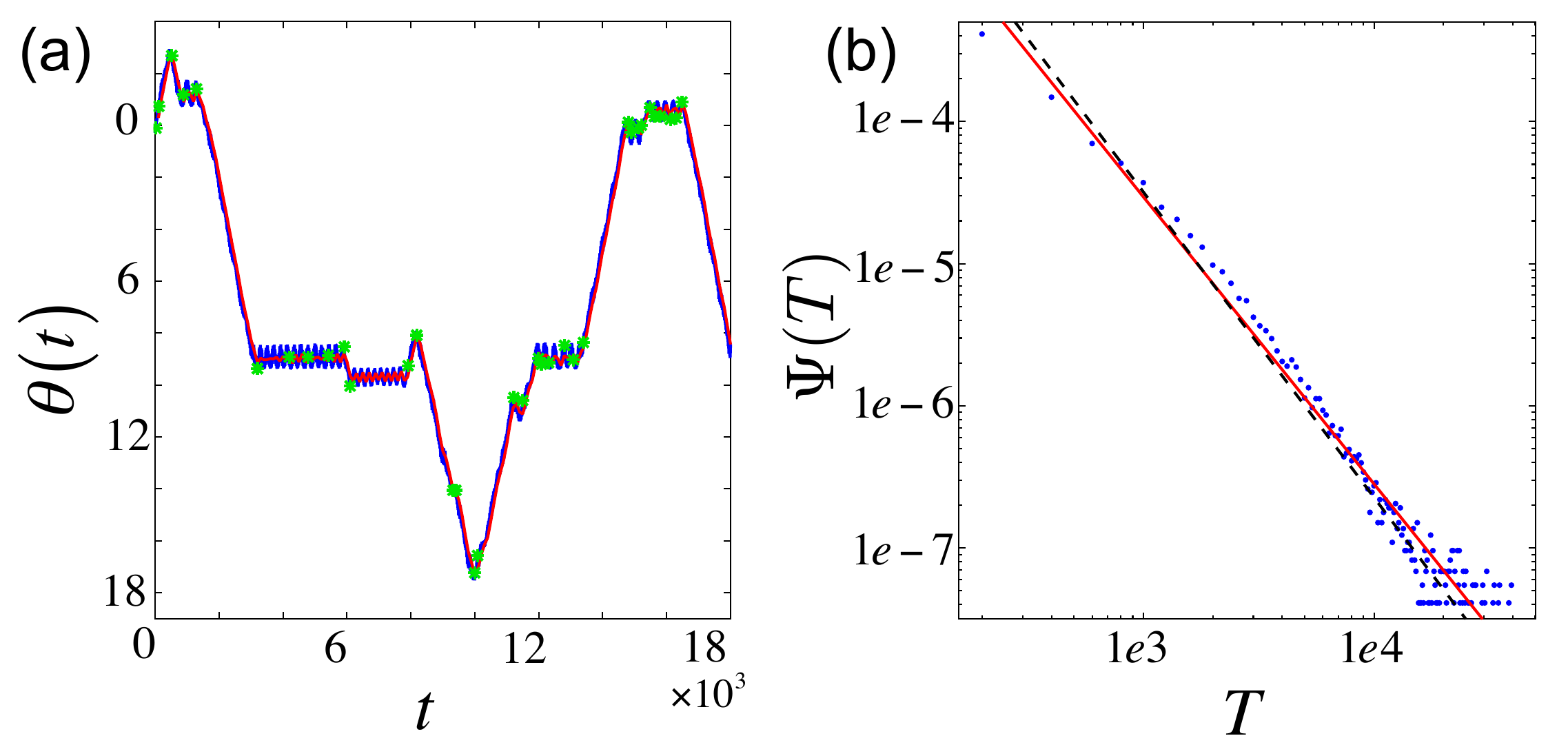}}
\caption{Calculating the PDF. (a) A piece of a L{\'{e}}vy-walk trajectory
is shown in blue. The sliding-window average of the trajectory overlaps
in magenta. In green asteri, the points identified by the algorithm
as starting points of new bouts are marked. The identification is
good for long bouts with a non-zero angular velocity, but marks too
many points in zero-velocity bouts. (b) The PDF $\Psi\left(T\right)$
vs. the bout lengths $T$ is presented in a log-log plot. In solid
red is the linear fit going through these points, while the dashed
black line is the slope predicted by the L{\'{e}}vy-walk model.}
\label{fig:PDFcalc}
\end{figure*}

The algorithm we used eventually is quite simple. First, it creates
a moving average of the trajectory, thus smoothing out the small oscillations
(Figure \ref{fig:PDFcalc}, red solid line). The next step is to divide the velocities
into negative, positive and zero, and identify the points at which
the velocity switches signs. Last, bouts that are shorter than the
average oscillation length are eliminated and we are left with the
final identification of the bouts (Figure \ref{fig:PDFcalc}(a), green asteri). 

In order to obtain good statistics, we take 10 trajectories of the
same ensemble, found by perturbing the initial conditions slightly.
We then calculate the PDF exponent of the bout lengths using the data
from all ten runs and fit it to a power law. However, as opposed to
the MSD calculation, this calculation doesn't grant us with a large
amount of statistics, rendering the error in $\nu$ very high. There
are several reasons for this. First of all, the algorithm has a problem
with the identification of the short bouts. Second of all, the tail
of the PDF, in which we are mainly interested, corresponds to very
long bouts which are quite rare, so a very long computation is required
to gain enough of these long bouts.

\subsection{Poincar{\'{e}} sections}
Visualizing the system's five dimensional phase space is a challenge primarily because of it high dimensionality. However, the phase space structure can be explored to some extent by a two-dimensional Poincar{\'{e}} map, constructed by plotting the points of intersection of certain trajectories with a set subspace of phase space to create a two-dimensional projection of a trajectory. In the plots presented in this appendix, many trajectories of the same energy differing by their initial conditions are plotted on the same graph to expose periodic orbits, fixed points and chaotic trajectories. The coordinates used in this visualisation are Birkhoff coordinates, a set of conjugate variables given by performing Birkhoff normal form expansion to the fourth order \cite{KE18}. They are related to the reduced variables by:
\begin{equation}
\begin{aligned}J_{1}=I_{1}\,,\,\,J_{2}=I_{1}+I_{2}\,,\,\,J_{3}=I_{3}\,\,,
\\
\,\,\psi_{1}=\varphi_{1}-\varphi_{2}\,,\,\,\psi_{2}=\varphi_{2}\,,\,\,\psi_{3}=\varphi_{3},\end{aligned}
\end{equation}
where $\left\{ I_{l},\varphi_{l}\right\} _{l=1}^{3}$ are the action-angle
coordinates of the linear system defined as $\vec{w}=(0,0,2)+\vec{\delta w}, \quad \vec{p}=\vec{\delta p}$, with the transformation:

\begin{equation}
I_{l}=\left(\frac{\delta w_{l}}{4\sqrt{2/\omega_{l}}}\right)^{2}+\left(\frac{\delta p_{l}}{\sqrt{\omega_{l}/8}}\right)^{2}\,,\,\varphi_{l}=\arctan\left(\frac{\omega_{l}}{16}\frac{\delta w_{l}}{\delta p_{l}}\right).
\end{equation}

The angle coordinate $\psi_1$, related to the phase difference between the two degenerate normal modes, corresponds directly to the rotational direction of the system; $0<\psi_1<\pi$ to clockwise rotation, $\pi<\psi_1<2\pi$ to counter-clockwise rotation. Also, from their construction $0\leq J_1\leq J_2$. Therefore, plotting a Poincar{\'{e}} map of $\psi_1$ vs. $J_1/J_2$ allows an identification between constant angular periodic trajectories, in which the trajectory stays in one of the two rectangles drawn by $0<J_1/J_2<1$ and either $0<\psi_1<\pi$ or $\pi<\psi_1<2\pi$, and L{\'{e}}vy-walk or chaotic trajectories in which the trajectory moves between these two rectangles. The Poincar{\'{e}} maps also reveal fixed points of the system: there are fixed points at $J_1/J_2=0.5$ and $\psi_1=\pi/2$ or $\psi_1=3\pi/2$ corresponding to a constant average angular velocity, and fixed points at $\psi_1=\pi$, $J_1/J_2=1$ and $J_1/J_2=0.25$ corresponding to zero average angular velocity. Around these fixed points, stable periodic trajectories persist for relatively high energies. They are studied analytically in \cite{KE18}.

At low energies, the trajectories show persistent periodicity around the various fixed points of the system, and a folliation of the phase space into separated periodic orbits is observed. As the energy rises, the periodic orbits begin to lose stability and intersect each other. At some point the periodicity breaks and trajectories begin migrating between fixed points, orbiting one fixed point for a certain amount of time before moving to orbit a different one. This migration between fixed points produces the so-called angular L{\'{e}}vy-walk motion. Also, the mixed nature of phase space is revealed, where at a given energy some trajectories remain regular while others have lost regularity, see Fig. 9(c,d). High enough energies break this pattern completely and the fixed points lose significance as all trajectories spread pretty much evenly around the section regardless of their starting point, perhaps pointing at ergodic motion.
%
%

\begin{figure*}[ht]
\centerline{\includegraphics[width=1\textwidth]{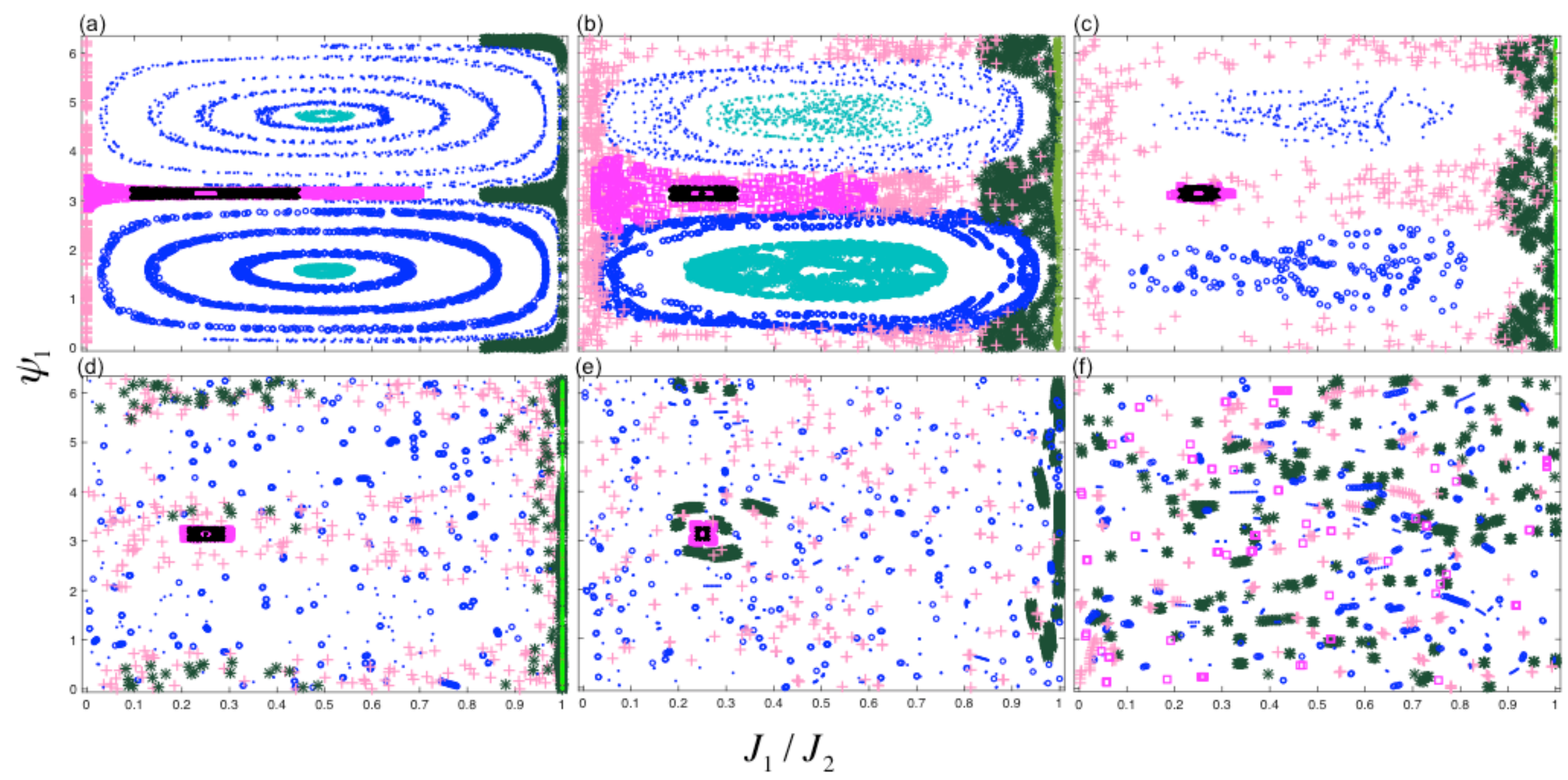}}
\caption{ Poincar{\'{e}} maps in Birkhoff coordinates (defined in appendix), demonstrate the transition to chaos with growing energy. Each subplot was created by plotting the intersection of different equal energy trajectories with a two dimensional surface of the five-dimensional equi-energy phase space. Different trajectories are well separated or marked differently. (a)  $E=0.0075$. All trajectories are regular and fixed points can be identified at $J_1/J_2=0.5,\psi_1=\pi/2$, $J_1/J_2=0.5,\psi_1=3\pi/2$, $J_1/J_2=0.25,\psi_1=\pi$ and $J_1/J_2=1,\psi_1=\pi$.  Also, the trajectory sticks to $J_1/J_2=0$. (b) $E=0.2193$. Trajectories are quasi-periodic, and the same fixed points can be identified, though the  $J_1/J_2=0$ line no longer traps trajectories. (c) $E=0.4$. Most trajectories are still quasi-periodic, but some exhibit L{\'{e}}vy-walk motion migrating between the top and bottom rectangles. (d)  $E=0.5605$. Most trajectories exhibit L{\'{e}}vy-walk motion, but the  $J_1/J_2=0.25,\psi_1=\pi$ fixed point still traps trajectories. (e) $E=0.7000$. Most trajectories are chaotic, travelling in a seemingly random manner around phase space. (f) $E=1.0000$. No fixed points remain; all trajectories seem chaotic and practically indistinguishable.}
\label{fig:Poincare}
\end{figure*}


\bibliographystyle{unsrt} 
\bibliography{refFinal2} 
\end{document}